\documentstyle[12pt]{article}
\begin{document}

\newcommand{\mod}[1]{\mbox{$\left| {#1} \right|$}}
\newcommand{\norm}[1]{\mbox{$\left\| {#1} \right\|$}}
\newcommand{\V}{\mbox{${\cal V}$}}

\vskip 2cm
\title{How to Create a 2-D Black Hole.}
\author{\\
V. Frolov${}^{*} {}^{1,2,3}$\,
S.Hendy${}^{\dag} {}^{1}$ and A.L.Larsen${}^{||} {}^{1}$}
\maketitle
\noindent
$^{1}${ \em
Theoretical Physics Institute, Department of Physics, \ University of
Alberta, Edmonton, Canada T6G 2J1}
\\ $^{2}${\em CIAR Cosmology Program}
\\ $^{3}${\em P.N.Lebedev Physics Institute,  Leninskii Prospect 53,
Moscow
117924, Russia}
\begin{abstract}
\baselineskip=1.5em

The interaction of a cosmic string with a four-dimensional stationary
black
hole is considered. If a part of  an infinitely long string passes
close to a
black hole it can be captured. The final stationary configurations of
such
captured strings are investigated. A uniqueness
theorem is proved, namely it is shown that the minimal 2-D surface
$\Sigma$ describing a captured stationary string coincides with a
{\it
principal Killing surface}, i.e. a surface formed by Killing
trajectories
passing through a principal null ray of the Kerr-Newman geometry.
Geometrical
properties of principal Killing surfaces are investigated and it is
shown that the internal geometry of $\Sigma$ coincides with the
geometry of  a
2-D black or white hole ({\it string hole}). The equations for
propagation of
string perturbations are shown to be identical with the equations for
a coupled
pair of scalar fields 'living' in the spacetime of  a 2-D string
hole. Some
interesting features of physics of 2-D string  holes are described.
In
particular, it is shown that the existence of the extra dimensions of
the
surrounding spacetime makes interaction possible between the interior
and
exterior of a string black hole; from the point of view of the 2-D
geometry
this interaction is acausal. Possible application of this result to
the
information loss puzzle is briefly discussed.

\end{abstract}


\noindent
$^{*}$Electronic address: frolov@phys.ualberta.ca\\
$^{\dag}$Electronic address: hendy@phys.ualberta.ca\\
$^{||}$Electronic address: alarsen@phys.ualberta.ca
\newpage
\baselineskip6.8mm
\section{Introduction}
\hspace{\parindent}
Black hole solutions in a  spacetime of lower than 4 dimensions have
been
discussed for a long time (see e.g. ref.\cite{brown} and references
therein).
Such solutions are of interest mainly because they provide toy models
which
allow one to investigate unsolved problems in four-dimensional black
hole
physics. The interest in 2-D black holes greatly increased after
Witten
\cite{Witt:91} and Mandal, Sengupta, and Wadia \cite{MaSeWa:91} have
shown that
2-D black hole solutions naturally arise in superstring motivated 2-D
dilaton
gravity. Many aspects of 2-D black hole physics and its relation to
4-D gravity
were discussed in a number of recent publications (see e.g.
ref.\cite{mann} ).
The main purpose of this paper is to show that there might exist
physical
objects which behave as 2-D black holes. Namely, we consider  a
cosmic string
interacting with a usual 4-D stationary black hole. If an infinitely
long
string passes close enough to the black hole it can be captured
\cite{moss,FroZel:89}. We study stationary final states of a captured
infinite
string, with endpoints fixed at infinity. We show that  there is only
a very
special family of solutions describing a stationary string which
enters the
ergosphere, namely the strings lying on cones of a given angle
$\theta=$const.
We demonstrate that the induced 2-D geometry of a stationary string
crossing
the  static limit surface and entering the ergosphere of a rotating
black hole
has the metric of a 2-D black or white hole.  The
horizon of such a 2-D string hole coincides with the intersection of
the string
world-sheet with the static limit surface. We shall also demonstrate
that the
2-D string hole geometry can be tested by studying the propagation of
string
perturbations. The perturbations propagating along the {\it cone
strings}
($\theta=$const.) are shown to obey the relativistic equations for a
coupled
system of two scalar fields. These results generalize the results of
ref.\cite{FrAl:95} where the corresponding equations were obtained
and
investigated for strings lying in the equatorial plane.  The quantum
radiation
of string excitations ({\em stringons}) and thermodynamical
properties of
string  holes are discussed. The remarkable property of 2-D string
holes as
physical objects is that  besides quanta (stringons)  living and
propagating
only on the 2-D world-sheet there exist other field quanta
(gravitons, photons
etc.) living and propagating in the surrounding physical 4-D
spacetime.  Such
quanta can enter the ergosphere as well as leave it and return back
to the
exterior. For this reason the presence of extra physical dimensions
makes
dynamical interaction possible between the interior and exterior of a
2-D
string black hole, which appears acausal from the perspective of the
internal
2-D geometry. The possible application of this effect to the
information loss
puzzle is briefly discussed.

The paper is organized as follows: In Section 2 we collect results
concerning
the  Kerr-Newman geometry which are necessary for the following
sections.  In
Section 3 we introduce the notion of a {\it principal Killing
surface} and we
prove that a principal Killing surface is a minimal 2-surface
embedded in the 4
dimensional spacetime. In Section 4 we prove the uniqueness theorem,
i.e. the
statement that the principal Killing surfaces are the only stationary
minimal
2-surfaces that are timelike and regular in the vicinity of the
static limit
surface of the Kerr-Newman black hole. In Section 4 we also relate
the
principal Killing surfaces with the world-sheets of a particular
class of
stationary cosmic strings -  the cone strings. In Section 5 we show
that the
internal  geometry of these world-sheets  is that of a
two-dimensional black or
white hole and we discuss the geometry of such string holes.  In
Section 6 we
consider
the propagation of perturbations along a stationary  string  using a
covariant
approach developed in ref.\cite{FrAl:94} (see also
refs.\cite{guv,car,vil}),
and we show that the corresponding equations coincide with a system
of coupled
equations for a pair of scalar fields on the two-dimensional string
hole
background. Finally in Section 7, we  discuss the physics of  string
holes and
give our conclusions.

\baselineskip6.8mm
\section{Kerr-Newman geometry}
\hspace{\parindent}

In Boyer-Lindquist coordinates the Kerr-Newman metric is given by
\cite{boy}:
\begin{equation}
\label{2.1}
ds^2=-\frac{\Delta}{\rho^2} [dt -a\sin^2\theta d\phi ]^2
+\frac{\sin^2\theta}{\rho^2}
[(r^2+a^2)d\phi -a dt]^2 +\frac{\rho^2}{\Delta}dr^2 +\rho^2 d\theta^2
,
\end{equation}
where  $\Delta=r^2-2Mr+Q^2+a^2$ and $\rho^2=r^2+a^2\cos^2\theta$.
The
corresponding electromagnetic field tensor is given by:
\begin{eqnarray}
{\bf F}\hspace*{-2mm}&=&\hspace*{-2mm}\frac{Q(r^2-
a^2\cos^2\theta)}{\rho^4} {\bf d}r\wedge[{\bf d}t-
a\sin^2\theta{\bf d}\phi]\nonumber\\
\hspace*{-2mm}&+&\hspace*{-2mm}\frac{2Qar}{\rho^4}\cos\theta\sin\theta
 {\bf
d}\theta\wedge[(r^2+a^2){\bf d}\phi-a{\bf d}t].
\end{eqnarray}

The
spacetime (2.1) possesses a Killing vector  $\xi^{\mu} = (1,0,0,0)$
which is
timelike
at infinity. The norm of the Killing vector is:
\begin{equation}
F\equiv-\xi^2 =1-\frac{2Mr-Q^2}{\rho^2}.
\end{equation}
A surface $S_{st}$ where $\xi$ becomes null $(F=0)$ is known as the
static
limit surface. It is defined by:
\begin{equation}
r=r_{st}\equiv M+\sqrt{M^2-Q^2-a^2\cos^2\theta}.
\end{equation}

The Kerr-Newman metric (\ref{2.1}) is of type $D$ and possesses two
principal
null directions $l^{\mu}_{+}$ and $l^{\mu}_{-}$. Each of these null
vectors
obey the relation:
\begin{equation}\label{3.1}
C^{(+)}_{\alpha\beta\gamma\delta}l^{\beta}l^{\delta}=
Cl_{\alpha}l_{\gamma} ,
\end{equation}
where:
\begin{equation}\label{3.2}
C^{(+)}_{\alpha\beta\gamma\delta}=C_{\alpha\beta\gamma\delta}+
iC^{*}_{\alpha\beta\gamma\delta},\hspace{1cm}
C^{*}_{\alpha\beta\gamma\delta}=1/2 e_{\alpha\beta\mu\nu}
C^{\mu\nu}_{\ \ \ \gamma\delta}.
\end{equation}
Here $C_{\alpha\beta\gamma\delta}$ is the Weyl tensor,
$e_{\alpha\beta\mu\nu}$
is the totally antisymmetric tensor, and $C_{\pm}$ are non-vanishing
complex
numbers. The Goldberg-Sachs theorem\cite{GoSa:62} implies that the
integral
lines $x^{\mu}(\lambda)$ of principal null directions
\begin{equation}\label{3.3}
{d{x^{\mu}}\over d\lambda_{\pm}}=\mp l^{\mu}_{\pm}
\end{equation}
are null geodesics $(l^{\mu}l_{\mu}=0,\;\;\;
l^{\mu} l^{\nu}_{\ ;\mu}=0)$ and their congruence is shear free. We
denote
by $\gamma_+$   and  $\gamma_-$  ingoing and outgoing principal null
geodesics,
respectively, and choose the parameter $\lambda_{\pm}$ to be  an
affine
parameter along the geodesic. The explicit form of $l_{\pm}$ is given
by:
\begin{equation}
\label{A1} l^{\mu}_{\pm}  =\left( (r^2+a^2)/\Delta, \mp1, 0, a/\Delta
\right),\;\;\;\;  \;\;\;\;\;  l_{\pm \mu} = \left( -1,\mp
\rho^2/\Delta, 0, a
\sin^2 \theta \right).
\end{equation}
The normalization has been chosen so that $l_{\pm}$ are future
directed and
such that:
\begin{equation}\label{3.4a}
l^{\mu}_+ l_{-\mu}=-2 \rho^2/ \Delta.
\end{equation}

The Killing equation implies that the tensor  $\xi_{\mu;\nu}$ is
antisymmetric
and  its eigenvectors with non-vanishing eigenvalues are null.  In
the
Kerr-Newman geometry $\xi_{\mu;\nu}$ is of the form:
\begin{equation}\label{3.7a}
\xi_{\mu;\nu}=(\Delta F^{\prime}/ 2 \rho^2)
l_{+[\mu}l_{-\nu]}+(2ia(1-F)\cos
\theta/\rho^2) m_{[\mu}\bar{m}_{\nu]},
\end{equation}
where we have made use of the complex null vectors $m$ and $\bar{m},$
that
complete the Kinnersley null tetrad. In the normalization where
$m^\mu\bar{m}_\mu=1,$ they take the form:
\begin{equation}
m^\mu=\frac{1}{\sqrt{2}\rho}(ia\sin\theta,0,1,i/\sin\theta),
\;\;\;\;\;\;m_\mu=\frac{1}{\sqrt{2}\rho}
(-ia\sin\theta,0,\rho^2,i(a^2+r^2)\sin\theta).
\end{equation}
The remarkable property of the Kerr-Newman geometry is that the
principal null
vectors $l_{\pm}$ are  eigenvectors of $\xi_{\mu;\nu}$. Namely one
has:
\begin{equation}\label{3.7b}
\xi_{\mu;\nu}l_{\pm}^{\nu}=\mp\kappa
l_{\pm\mu},\hspace{1cm}\kappa=\pm{1\over
2}l^{\nu}_{\pm}(\xi^2)_{;\nu}=\frac{1}{2}F_{,r}
=\frac{Mr^2-rQ^2-Ma^2\cos^2\theta}{\rho^4}.
\end{equation}
These equations, (\ref{3.7a}) and (\ref{3.7b}),  will play an
important role
later in our analysis.

Notice also that the electromagnetic field tensor ${\bf F}$ has the
form:
\begin{equation}
F_{\mu\nu}=-\frac{\Delta}{\rho^2}
\left(\frac{Q(r^2-a^2\cos^2\theta)}
{\rho^4}\right)l_{+[\mu}l_{-\nu]}+
\frac{4iQar\cos\theta}{\rho^4}m_{[\mu}\bar{m}_{\nu]},
\end{equation}
so that:
\begin{equation}
F_{\mu\nu}l^\nu_{\pm}=\mp\frac{Q(r^2-a^2\cos^2\theta)}
{\rho^4}l_{\mu\pm}.
\end{equation}

\baselineskip6.8mm
\section{Principal Killing surfaces}
\hspace{\parindent}
Our aim is to consider stationary configurations of cosmic strings in
the
gravitational field of a charged rotating black hole. In particular,
we are
interested in the situation when a string is trapped by a black hole;
that is
when the string crosses the black holes static limit surface and
enters the
ergosphere.  We neglect the thickness of the string and its own
gravitational
field. In this approximation the string evolution is described by a
timelike
2-D world-sheet (for general properties of cosmic strings, see for
instance
refs.\cite{vil2,shell}). The dynamical equations obtained by
variation of the
Nambu-Goto action for a string imply that this world-sheet is a
minimal
surface. So the mathematical problem we are trying to solve is to
find
stationary timelike minimal surfaces which intersect the static limit
surface
of a rotating black hole.
For this purpose we begin by considering the general properties of
stationary
timelike surfaces.

Let $\Sigma$ be a two-dimensional timelike surface embedded in a
stationary
spacetime, and let $\xi$ be the corresponding Killing vector  which
is timelike
at infinity. $\Sigma$ is said to be {\it stationary} if it is
everywhere
tangent to the Killing vector field $\xi$.  For any such surface
$\Sigma$ there
exists two  linearly independent null vector fields $l,$ tangent to
$\Sigma.$
We assume that the integral curves of $l$ form a congruence and cover
$\Sigma$
(i.e. each point $p \in \Sigma$ lies on exactly one of these integral
curves).

Thus we can construct a stationary timelike surface $\Sigma$ in the
following
way: consider a null ray $\gamma$ with tangent vector field $l$  such
that
$\xi \cdot l$ is non-vanishing everywhere along $\gamma$. There is
precisely
one Killing trajectory with tangent vector $\xi$ that passes through
each point
$p \in \gamma$. This set of Killing trajectories passing through
$\gamma$ forms
a stationary 2-D surface $\Sigma$. We define $l$ over $\Sigma$ by Lie
propagation along each Killing trajectory. We call $\gamma$ a basic
ray of
$\Sigma.$ It is easily verified that $l$ remains null when defined in
this
manner over $\Sigma$.

We can use the Killing time parameter $u$ and the affine parameter
$\lambda$
along $\gamma$ as coordinates on $\Sigma.$ In these coordinates
$\zeta^A=(u,\lambda)$ one has $x^{\mu}_{,0}=\xi^{\mu}$ and
$x^{\mu}_{,1}=
l^{\mu}$ and the induced metric $G_{AB}=g_{\mu \nu} x^{\mu}_{,A}
x^{\nu}_{,B}\;\;(A,B,...=0,1)$ is of the form:

\begin{equation}\label{5.1}
dS^2 =G_{AB}d\zeta^A d\zeta^B = - F du^2 + 2(\xi \cdot l) du
d\lambda.
\end{equation}
In the case of a black hole  the Killing vector $\xi$ becomes null at
the
static limit surface $S_{st}. $ In what follows we always choose $l$
to be that
of two possible null vector fields
 on $\Sigma$ which does not coincide with $\xi$ on the static limit
surface
$S_{st}$.  In this case the metric (\ref{5.1}) is regular at
$S_{st}.$
Now introduce two vectors $n_R^{\mu}$  (R=2,3) normal to the 2-D
surface
$\Sigma$:
\begin{equation}
\label{4.1}g_{\mu \nu} n_R^{\mu} n_S^{\nu} = \delta_{RS},
\;\;\;\;\;\;\;\;\;\;
g_{\mu \nu} x^{\mu}_{,A} n_R^{\nu} = 0,
\end{equation}
which satisfy the completeness relation:
\begin{equation}
\label{4.2}g^{\mu \nu} = G^{AB} x^\mu_{,A} x^\nu_{,B} +  \delta^{RS}
n_R^{\mu}
n_S^{\nu}.
\end{equation}
These two normal vectors span the vector space normal to the surface
at a given
point, and they are uniquely defined up to local rotations in the
$(n_2,n_3)$-plane.

The  second fundamental form is defined as:
\begin{equation}
\label{4.4a}
\Omega_{RAB}  =  g_{\mu\nu}n^\mu_R x^\rho_{,A}\nabla_\rho x^\nu_{,B}.
\end{equation}
The condition that a surface $\Sigma$ is minimal can be written in
terms of the
trace of the second fundamental form as follows:

\begin{equation}\label{5.2}
\Omega_{RA}\hspace*{1mm}^{A} \equiv G^{AB}\Omega_{RAB}=0.
\end{equation}
We find that in the metric (\ref{5.1}) the second fundamental form is
given by:
\begin{eqnarray}\label{5.3}
\nonumber \Omega_{RA}\hspace*{1mm}^{A}  &=& g_{\mu \nu} G^{AB}
n_R^{\mu}
x^{\gamma}_{,A} \nabla_{\gamma} x^{\nu}_{,B} \\
&=& g_{\mu \nu} n_R^{\mu} ({2 \over (\xi \cdot l)} l^{\gamma}
\nabla_{\gamma}
\xi^{\nu} + {F \over (\xi \cdot l)^2} l^{\gamma} \nabla_{\gamma}
l^{\nu}).
\end{eqnarray}

Consider a special type of a stationary timelike 2-surface in the
Kerr-Newman
geometry. Namely  a surface for which  the null vector $l$ coincides
with one
of the principal null geodesics $l_\pm$ of the Kerr-Newman geometry.
We call
such surface $\Sigma_{\pm}$ a {\it  principal Killing surface} and
$\gamma_{\pm}$ its {\it basic ray}. We shall use indices $\pm$ to
distinguish
between quantities connected with $\Sigma_{\pm}$.  The fact that
$l_{\pm}$ are
geodesics ensures that $ l^{\gamma}_{\pm} \nabla_{\gamma}
l^{\mu}_{\pm}\propto
l^{\mu}_{\pm}.$  In addition, from equation (\ref{3.7b}),
$l^{\gamma}_{\pm}
\nabla_{\gamma} \xi^{\mu}\propto l^{\mu}_{\pm}$ which, because of the
contraction with $n^\nu_R,$ guarantees that $
\Omega_{RA}\hspace*{1mm}^{A}$
vanishes for a principal Killing surface, i.e. every principal
Killing surface
is minimal. Thus $\Sigma_{\pm}$ are stationary solutions of the
Nambu-Goto
equations.

It should be stressed that the principal Killing surfaces are only
very special
stationary minimal surfaces. A principal Killing surface is uniquely
determined
by indicating two coordinates (angles) of a point where it crosses
the static
limit surface. Because of the axial symmetry only one of these two
parameters
is non-trivial. A general stationary string solution in the
Kerr-Newman
spacetime can be obtained by separation of variables
(ref.\cite{FroZel:89}, see
also Section 4) and it depends on 3 parameters (2 of which are
non-trivial).

\baselineskip6.8mm
\section{Uniqueness Theorem}
\hspace{\parindent}
We  prove now that the only stationary timelike minimal 2-surfaces
that cross the static limit surface $S_{st}$ and are regular in its
vicinity
are the principal Killing surfaces.

Consider a stationary timelike surface $\Sigma$ described by the line
element
(3.1).
By using the completeness relation (3.3) and the metric (3.1) we
obtain:
\begin{equation}\label{5.4a}
\Omega^2 =  z^{\mu} z_{\mu},\;\;\;\;\;\;\;\;\;\;z^{\mu}
\equiv\frac{2}{(\xi\cdot l)}l^{\gamma} \nabla_{\gamma} \xi^{\mu} + {F
\over
(\xi \cdot l)^2} l^{\gamma} \nabla_{\gamma} l^{\nu}.
\end{equation}
In other words $\Sigma$ is minimal if and only if  $z^\mu$ is null so
that
$\Omega^2=0$ (clearly if $\Sigma$ is a principal Killing surface then
$z^{\mu}
\propto l_{\pm}^\mu$ and this condition is satisfied). In general we
observe
that $l \cdot z$ vanishes as $l^\mu$ is null and as $\xi_{\mu;\nu}$
is
antisymmetric. Thus if $z^\mu$ is null then it must be proportional
with
$l^\mu$.
The condition that $\Omega^2=0$ in the line element (3.1) then
becomes:

\begin{equation}
\label{5.5}
2 (\xi \cdot l) l^\rho\nabla_\rho\xi^\mu+Fl^\rho\nabla_\rho
l^\mu+(\xi \cdot l)
l^\rho\frac{d}{dx^\rho} \left( {F \over \xi \cdot l}\right)l^\mu=0.
\end{equation}
It is easily verifed that equation (\ref{5.5}) is invariant under
reparametrizations of $l^{\mu},$ i.e. if $l^\mu$ satisfies
(\ref{5.5}) then so
does $g(x) l^\mu$. Thus without loss of generality we may normalize
$l^\mu$ so
that $l \cdot \xi = -1$. Then (\ref{5.5}) becomes:
\begin{equation}
-2l^\rho\nabla_\rho\xi^\mu+Fl^\rho\nabla_\rho
l^\mu+l^\rho\frac{dF}{dx^\rho}l^\mu=0.
\end{equation}
Since $l^\rho\nabla_\rho l^\mu$ is regular on $\Sigma$, this equation
at the
static limit surface ($F=0$) reduces to:
\begin{equation}
(\xi_{\mu;\rho}-\frac{1}{2}\frac{dF}{dx^\rho}l_\mu)l^\rho=0,
\end{equation}
that is, $l^\rho$ is a real eigenvector of $\xi_{\mu;\rho}.$ From
equation
(2.10) follows that the only real eigenvector of $\xi_{\mu;\rho}$
must be
either
$l_+$ or $l_-.$ Thus we have $l\propto l_\pm$ at the static limit
surface.

Now suppose there exists a timelike minimal surface $\Sigma$
different from
$\Sigma_\pm.$ At the static limit surface $\Sigma$ must have
$l\propto l_+\;$
(or $l\propto l_-$). In the vicinity of the static limit surface, $l$
can have
only small deviations from $l_+.$
 From the conditions $l\cdot l=0$ and $l\cdot \xi=-1,$ we then get
the
following general form of $l$ in the vicinity of the static limit
surface:
\begin{equation}
l=[1+\frac{ia\sin\theta}{\sqrt{2}\rho}(B-\bar{B})]
l_++\bar{B}m+B\bar{m}+{\cal
O}(B^2),
\end{equation}
up to first order in $(B,\bar{B}).$ We then insert this expression
into (4.2),
contract by $\bar{m}_\mu$ and keep only terms linear in
$(B,\bar{B}):$
\begin{equation}
-2\bar{m}_\mu
l^\rho\nabla_\rho\xi^\mu=\frac{-2ia(1-F)\cos\theta}
{\rho^2}\bar{B}+{\cal O}(B^2),
\end{equation}
\begin{equation}
\bar{m}_\mu
l^\rho\frac{dF}{dx^\rho}l^\mu=l^\rho_+\frac{dF}{dx^\rho}\bar{B}+
{\cal
O}(B^2)=-F'\bar{B}+{\cal O}(B^2),
\end{equation}
\begin{eqnarray}
\bar{m}_\mu F\l^\rho\nabla_\rho
l^\mu\hspace*{-2mm}&=&\hspace*{-2mm}Fl_+^\rho\frac{d\bar{B}}{dx^\rho}-
2Fl_+^\mu
m^\rho\bar{m}_{(\rho;\mu)}\bar{B}+F\bar{m}^\rho\bar{m}_\mu
l^\mu_{+;\rho}
B+{\cal O}(B^2)\nonumber\\
\hspace*{-2mm}&=&\hspace*{-2mm}-F\frac{d\bar{B}}{dr}-
\frac{(\rho+2ia\cos\theta)F}{\rho^2}\bar{B}+{\cal O}(B^2),
\end{eqnarray}
where the last equality was obtained by direct calculation using
(2.8),
(2.11).
Thus altogether:
\begin{equation}
F\frac{d\bar{B}}{dr}=-\bar{B}[\frac{dF}{dr}+\frac{2ia\cos\theta}
{\rho^2}+
\frac{F}{\rho}]+{\cal O}(B^2).
\end{equation}
It is convenient to rewrite this equation in the form:
\begin{equation}
\frac{d\bar{B}}{dr^*}=-\Omega\bar{B};\;\;\;\;\;\;\;\;\;\;\Omega\equiv
\frac{dF}{dr}+\frac{2ia\cos\theta}{\rho^2}+\frac{F}{\rho}
\end{equation}
and we have introduced the tortoise-coordinate $r^*$ defined by:
\begin{equation}
\frac{dr}{dr^*}=F(r).
\end{equation}
Near the static limit surface the complex frequency $\Omega$ is given
by:
\begin{equation}
\Omega=\frac{2(r_{st}-M+ia\cos\theta)}
{r_{st}^2+a^2\cos^2\theta}+{\cal
O}(r-r_{st})\equiv\Omega_{st}+{\cal O}(r-r_{st})
\end{equation}
The solution of equation (4.10) near the static limit surface is then
given by:
\begin{equation}
\bar{B}=c e^{-\Omega_{st}r^*};\;\;\;\;\;\;\;\;\;\;c=\mbox{const.}
\end{equation}
Notice that $Re(\Omega_{st})>0,$ thus $\bar{B}$ is oscillating with
infinitely
growing amplitude near the static limit surface.
 A solution regular near the static limit surface $(r^*\rightarrow
-\infty)$
can therefore only be obtained for $c=0,$ which implies that
$B=\bar{B}=0,$
thus we have shown that $\Sigma$ is minimal if and only if
 $l \propto l_\pm$. This proves the uniqueness theorem:
The only stationary timelike minimal 2-surfaces
that cross the static limit surface $S_{st}$ and are regular in its
vicinity
are the principal Killing surfaces.

We now discuss the physical meaning of this result.
For that purpose
it is convenient to introduce the ingoing ($+$) and outgoing ($-$)
Eddington-Finkelstein coordinates $(u_{\pm},\varphi_{\pm}):$

\begin{equation}
\label{2.12}
du_{\pm}=dt \pm \Delta^{-1} (r^2+a^2)dr,\hspace{1cm}
d\varphi_{\pm}=d\phi \pm \Delta^{-1}a dr ,
\end{equation}
and  to rewrite the Boyer-Lindquist metric (\ref{2.1}) as:
\begin{eqnarray}
\label{2.13}
ds^2\hspace*{-2mm}&=&\hspace*{-2mm}
-\frac{\Delta}{\rho^2}[du_{\pm} -a\sin^2\theta d\varphi_{\pm}]^2
+\frac{\sin^2\theta}{\rho^2}[(r^2+a^2)d\varphi_{\pm} -a
du_{\pm}]^2\nonumber \\
\hspace*{-2mm}&+&\hspace*{-2mm} \rho^2 d\theta^2 \pm 2dr [du_{\pm}
-a\sin^2\theta d\varphi_{\pm} ].
\end{eqnarray}
The electromagnetic field tensor (2.2) is:
\begin{eqnarray}
{\bf F}\hspace*{-2mm}&=&\hspace*{-2mm}\frac{Q(r^2-
a^2\cos^2\theta)}{\rho^4}{\bf d}r\wedge[{\bf d}u_\pm-
a\sin^2\theta{\bf d}\varphi_\pm]\nonumber\\
\hspace*{-2mm}&+&\hspace*{-2mm}\frac{2arQ\cos\theta\sin
\theta}{\rho^4} {\bf
d}\theta\wedge[(r^2+a^2){\bf d}\varphi_\pm-a{\bf d}u_\pm].
\end{eqnarray}

We have shown that any stationary minimal 2-surface that crosses the
static
limit must have $x^{\mu}_{,1} = l^\mu_{\pm}$ (up to a constant
factor). Using
the explcit form of $l_{\pm}$ in Boyer-Lindquist coordinates
(\ref{A1}) we can
choose the affine parameter along $\gamma_\pm$ to coincide with $r$
such that
$x'=\mp l_\pm,$ where the prime denotes derivative with respect to
$r.$ We can
then read off $\theta^{\prime}$ and $\phi^{\prime}$ for these
surfaces
$\Sigma_\pm$:
\begin{equation}\label{5.7}
\theta^{\prime} = 0, \;\;\;\;\;\;\;\; \varphi_\pm = \mbox {const.}
\end{equation}
In the Eddington-Finkelstein coordinates the induced metric on
$\Sigma_\pm$ is
then:
\begin{equation}
\label{2.15}
dS^2=-Fdu_{\pm}^2 \pm 2dr
du_{\pm};\;\;\;\;\;\;\;\;F=1-\frac{2Mr-Q^2}{\rho^2}.
\end{equation}
The induced electromagnetic field tensor is:
\begin{equation}
{\bf F}=\frac{Q}{\rho^4}(r^2-
a^2\cos^2\theta){\bf d}r\wedge{\bf d}u_\pm
\end{equation}
that is, the induced electric field is:
\begin{equation}
E_{r}=\frac{Q}{\rho^4}(r^2-
a^2\cos^2\theta).
\end{equation}

Equations (4.17) imply that a principal Killing string is located at
the cone
surface $\theta=$const.
These so called {\it cone strings} are thus the only stationary
world-sheets
that can
cross the static limit surface and are timelike and regular in its
vicinity. It
was shown, on the other hand, that the general stationary string
solution in
the Kerr-Newman spacetime can be obtained by separation of
variables\cite{FroZel:89}:
\begin{eqnarray}
(H_{rr}\frac{dr}{d\lambda})^2\hspace*{-2mm}&=&\hspace*{-2mm}
\frac{a^2b^2}{\Delta^2}-\frac{q^2}{\Delta}+1,\nonumber\\
(H_{\theta\theta}\frac{d\theta}{d\lambda})^2\hspace*{-2mm}
&=&\hspace*{-2mm}q^2-\frac{b^2}{\sin^2\theta}-a^2\sin^2\theta,\\
(H_{\phi\phi}\frac{d\phi}{d\lambda}
)^2\hspace*{-2mm}&=&\hspace*{-2mm}b^2,\nonumber
\end{eqnarray}
where $b$ and $q$ are arbitrary constants, while:
\begin{equation}
H_{rr}=\frac{\Delta-a^2\sin^2\theta}{\Delta},
\;\;\;\;\;\;H_{\theta\theta}=
\Delta-a^2\sin^2\theta,\;\;\;\;\;\;H_{\phi\phi}=
\Delta\sin^2\theta.
\end{equation}
In this general three-parameter family of solutions, parametrized by
$b,q$  and
some initial angle $\phi_0$, the stationary strings crossing the
static limit
surface are determined by (4.17), that is:
\begin{equation}
\varphi_\pm=\mbox{const.},\;\;\;\;\;\;\;\;q^2=2ab,\;\;\;\;\;\;\;\;\
\sin^2\theta={\mbox{const.}}=b/a,
\end{equation}
i.e. a two-parameter family of solutions (notice however that due to
the axial
symmetry only one of these parameters $b$ is non-trivial).
Physically it means
that a stationary cosmic string can only enter the ergosphere in very
special ways, corresponding to the angles (4.23).

\baselineskip6.8mm
\section{Geometry of 2-D string  holes}
\hspace{\parindent}

The metric (4.18) for $\Sigma_+$ describes a black hole, while  for
$\Sigma_-$
it describes a white hole. For $a=0,$ $\Sigma_\pm$ are geodesic
surfaces in the
4-D spacetime and they describe two branches of a geodesically
complete 2-D
manifold. However, it should be stressed that for the generic
Kerr-Newman
geometry $(a\neq 0)$, only one of two null basic lines of the
principal Killing
surface, namely the ray $\gamma_{\pm}$ with tangent vector $l{\pm}$,
is
geodesic in the four-dimensional embedding space. The other basic
null ray is
geodesic in $\Sigma_\pm$ but not in the embedding space. This implies
that in
general (when $a\neq 0$) the principal Killing surface is not
geodesic.
Furthermore, it can be shown that $\Sigma_{\pm}$ considered as a 2-D
manifold
is geodesically incomplete with respect to its null geodesic
$\gamma'$.

As a consequence of $\Sigma_\pm$ not being geodesic (when $a\neq 0$),
it is
possible, as we shall now demonstrate,  to send causal signals from
the inside
of the 2-D black hole to the outside of the 2-D black hole by
exploiting the 2
extra dimensions of the 4-D spacetime.

It is evident that there exist causal lines leaving the ergosphere
and entering
the black hole exterior. It means that "interior" and "exterior" of a
2-D black
hole can be connected by 4-D causal lines. We show now that (at least
for the
points lying close to the static limit surface) the causal line can
be chosen
as a null geodesic.
Consider for simplicity the stationary string corresponding to
($\theta=\pi/2,\;\varphi_+=0$) and crossing the static limit surface
in the
equatorial plane of a Kerr black hole. We will
demonstrate that there exists an outgoing null geodesic in the 4-D
spacetime
connecting the point $(r,\varphi_+)=(2M-\epsilon,0)$ of the cosmic
string
inside the ergosphere with the point $(r,\varphi_+)=(2M+\epsilon,0)$
of the
cosmic string outside the ergosphere, for $\epsilon$ small. An
outgoing null
geodesic, corresponding to positive energy at infinity $E$ and
angular momentum
at infinity $L_z$ in the equatorial plane of the Kerr black hole
background, is
determined by\cite{mtw}:
\begin{equation}
r^2\frac{dr}{d\lambda}={\cal P},
\end{equation}
\begin{equation}
r^2\frac{du_+}{d\lambda}=-a{\cal U}+\frac{r^2+a^2}{\Delta}[{\cal
P}+{\cal Q}],
\end{equation}
\begin{equation}
r^2\frac{d\varphi_+}{d\lambda}=-{\cal U}+\frac{a}{\Delta}[{\cal
P}+{\cal Q}],
\end{equation}
where:
\begin{equation}
{\cal U}\equiv aE-L_z,\;\;\;\;\;\;{\cal Q}\equiv Er^2+a{\cal
U},\;\;\;\;\;\;{\cal P}^2\equiv{\cal Q}^2-\Delta{\cal U}^2.
\end{equation}
and we consider the case where $dr/d\lambda>0.$
Inside the ergosphere the 4-D geodesic must follow the rotation of
the black
hole because of the dragging effect, that is, $d\varphi_+/d\lambda>0$
(for
$a>0$). However, after leaving the ergosphere the geodesic can reach
a turning
point in $\varphi_+$ and then return $(d\varphi_+/d\lambda<0)$
towards the
cosmic string outside the static limit surface. To be  more precise:
provided
$-L_z>aE,$ there will be a turning point in $\varphi_+$ outside
the static limit surface at $r=r_0$:
\begin{equation}
r_0=\frac{2M(aE-L_z)}{-L_z-aE}>2M.
\end{equation}
Obviously the turning point in $\varphi_+$ can be put at any value of
$r$
outside the static limit surface.
If we choose $E$ and $L_z$ such that:
\begin{equation}
r_0=2M+\epsilon-\frac{M}{2a^2}\epsilon^2,
\end{equation}
then, after reaching the turning point in $\varphi_+,$  the geodesic
will
continue in the direction opposite to the rotation of the 4-D black
hole with
constant $r=2M+\epsilon$ (to first order in $\epsilon$) and
eventually reach the point $(r,\varphi_+)=(2M+\epsilon,0)$ of the
cosmic string
outside the ergosphere.
\vskip 12pt
\hspace*{-6mm}We close this section with the following remarks:

Notice that the (outer) horizon of the 2-D black hole coincides with
the static
limit of the 4-D rotating black hole. The 2-D surface gravity, which
is
proportional to the 2-D temperature, is given by:
\begin{equation}
\kappa^{(2)}=\left.
\frac{1}{2}\frac{dF}{dr}\right|_{r=r_{st}}=
\frac{\sqrt{M^2-Q^2-a^2\cos^2\theta}}
{2M^2-Q^2+2M\sqrt{M^2-Q^2-a^2\cos^2\theta}}.
\end{equation}
The surface gravity of the 4-D Kerr-Newman black hole is:
\begin{equation}
\kappa^{(4)}=\frac{\sqrt{M^2-Q^2-a^2}}
{2M^2-Q^2+2M\sqrt{M^2-Q^2-a^2}},
\end{equation}
and then it can be easily shown  that:
\begin{equation}
\kappa^{(2)}\geq\kappa^{(4)}.
\end{equation}
That is to say, the 2-D temperature is higher than the 4-D
temperature (except
at the poles where they coincide) and it is always positive. Even if
the 4-D
black hole is extreme, the 2-D temperature is non-zero.

As we show in Appendix A, the  solutions of the form (4.18) can also
be
obtained in 2-D dilaton gravity:
\begin{equation}
{\cal S}=\frac{1}{2\pi}\int dt
dx\;\sqrt{-g}\;e^{-2\phi}\;[R+2(\nabla\phi)^2+V(\phi)],
\end{equation}
with the following dilaton potential:
\begin{equation}
V(\phi)=[\frac{2}{r^2}(rF)_{,r}]_{|r=e^{-\phi}/\lambda},
\end{equation}
if the dilaton field has the form:
\begin{equation}
\phi=-\log(\lambda r),\;\;\;\;\;\;\;\;\;\;\;\;\lambda=\mbox{const.}
\end{equation}
It should be stressed that this observation does not mean that we can
use the
dilaton-gravity equations in order to describe the dynamics of 2-D
string
holes, or to determine the back reaction of the string excitations on
the
geometry of string holes.

\baselineskip6.8mm
\section{String perturbation propagation}
\hspace{\parindent}

A general transverse perturbation about a background Nambu-Goto
string
world-sheet can be written as (summing over the $R$ indices):
\begin{equation}
\label{4.3}
\delta x^{\mu} = \Phi^R n_R^{\mu},
\end{equation}
where the normal vectors are defined by equations (3.2).
The equations of motion for the perturbations, $\Phi_R$ follow from
the
following effective action for stringons\cite{FrAl:94}:
\begin{equation}
\label{4.5}
{\cal S}_{\mbox{eff.}}=\int_{}^{}
d^2\zeta\sqrt{-G}\;\Phi^R\left\{G^{AB}
(\delta^T_R\nabla_A+\mu_R\hspace*{1mm}^T\hspace*{1mm}_A)
(\delta_{TS}\nabla_B+\mu_{TSB})+
\V_{RS}\right\}\Phi^S,
\end{equation}
where $\V_{RS} = \V_{(RS)}$ are scalar potentials and
$\mu_{RSA}=\mu_{[RS]A}$
are vector potentials which coincide with the normal fundamental
form:
\begin{equation}
\mu_{RSA}  =  g_{\mu\nu}n^\mu_R x^\rho_{,A}\nabla_\rho n^\nu_S.
\end{equation}
The scalar potentials
are defined as:
\begin{equation}
\label{4.6}
\V_{RS}\equiv\Omega_{RAB}\Omega_S\hspace*{1mm}^{AB}-
G^{AB}x^\mu_{,A}x^\nu_{,B}R_{\mu\rho\sigma\nu}n^\rho_R
n^\sigma_S.
\end{equation}
The equations describing the propagation of perturbations on the
world-sheet
background
are then found to be:
\begin{equation}
\label{4.7}
\left\{ \delta_{RS} \hspace*{1mm} {\,\lower0.9pt\vbox{\hrule
\hbox{\vrule
height 0.3 cm \hskip 0.3 cm \vrule height 0.3 cm}\hrule}\,} + 2
\mu_{RS}\hspace*{1mm}^A \partial_A  +  \nabla_A
\mu_{RS}\hspace*{1mm}^A -
\mu_{R} \hspace*{1mm}^{TA} \mu_{STA} + \V_{RS} \right\} \Phi^S = 0.
\end{equation}

We note that the perturbations (\ref{4.3}) and the effective action
(\ref{4.5})
are invariant under rotations of the normal vectors i.e. invariant
under the
transformations $n_R \mapsto \tilde{n}_R = \Lambda_R \;^S
n_S,\;\;\Phi^R\rightarrow\tilde{\Phi}^R=\Lambda^R\;_S\Phi^S,$ where:
\begin{equation}\label{4.7a}
\left[ \; \Lambda \; \right]_R\;^S = \pmatrix{
\cos \Psi & -\sin \Psi \cr
\sin \Psi & \cos \Psi \cr
},
\end{equation}
for some arbitrary real function $\Psi$. Thus we have a 'gauge'
freedom in our
choice of normal vectors.

Consider the scalar potential $\V_{RS} \equiv\Omega_{RAB}
\Omega_S\hspace*{1mm}^{AB}-
G^{AB}x^\mu_{,A}x^\nu_{,B}R_{\mu\rho\sigma\nu}n^\rho_R
n^\sigma_S$. It is easily verified that the first term
$\Omega_{RAB}\Omega_S\hspace*{1mm}^{AB}$ vanishes for the principal
Killing
surface $\Sigma_{\pm}$ independently of any choice of normal vectors
$n^\rho_R$. It is also possible to show that
the second term on the right hand side is invariant under rotations
of the
vectors $n_R,$ i.e. gauge invariant, in the Kerr-Newman spacetime
(see Appendix
\ref{B}). The symmetry and gauge invariance of  $\V_{RS}$ show that
it must be
proportional to $\delta_{RS}$ i.e. $\V_{RS} = \V \; \delta_{RS}$.
Now, using the completeness relation (\ref{4.2}) we find:
\begin{eqnarray}\label{4.8}
\nonumber \V & = & 1/2 \; \delta^{RS} \V_{RS} \\
\nonumber & = & - 1/2 \;
G^{AB}x^\mu_{,A}x^\nu_{,B}R_{\mu\rho\sigma\nu}
\delta^{RS} n_R^{\rho} n_S^{\sigma} \\
& = &1/2 \; G^{AB}x^\mu_{,A}x^\nu_{,B} \left(R_{\mu \nu} +
G^{CD}x^\rho_{,C}x^\sigma_{,D}R_{\mu\rho\sigma\nu} \right).
\end{eqnarray}

Making use of a representation of the Ricci tensor $R_{\mu \nu}$ in
terms of
the Kinnersley null tetrad, namely:
\begin{equation}\label{4.9}
R_{\mu \nu} = {2 Q^2 \over \rho^4} \left(m_{(\mu}  \bar{m}_{\nu)} +
(\Delta/ 2
\rho^2) l_{+(\mu}l_{-\nu)}\right),
\end{equation}
we are able to calculate the first term of equation (6.7) as follows:
\begin{equation}\label{4.10}
G^{AB}x^\mu_{,A}x^\nu_{,B} R_{\mu \nu} =  \pm 2 (\mp R_{\mu \nu}
l_{\pm}^{\mu}
\xi^{\nu})
=  {2 Q^2 \over \rho^4}  l_{\pm}^{\mu} \xi_{\mu}
=  - {2 Q^2 \over \rho^4}.
\end{equation}
To calculate the second term of equation (6.7) we use the
Gauss-Codazzi
equations \cite{eis} for a 2-surface $\Sigma$ embedded in a
4-dimensional
spacetime. Namely:

\begin{equation}\label{4.11}
R^{(2)}_{ABCD} = \left(\Omega_{RAC} \Omega^R\hspace*{1mm}_{BD} -
\Omega_{RAD}
\Omega^R\hspace*{1mm}_{BC}\right) + R_{\mu\rho\sigma\nu}
x^\mu_{,A}x^\rho_{,B}
x^\sigma_{,C}x^\nu_{,D}.
\end{equation}
Contracting (\ref{4.11}) over $A$ and $C$, and then $B$ and $D$ one
finds that
the scalar curvature on $\Sigma$ is just the sectional curvature in
the tangent
plane of $\Sigma$ i.e.:
\begin{equation}\label{4.12a}
R^{(2)} = G^{AC} G^{BD} R_{\mu\rho\sigma\nu} x^\mu_{,A}x^\rho_{,B}
x^\sigma_{,C}x^\nu_{,D}
\end{equation}
which is identically the second term in equation (\ref{4.8}), except
for the
sign. Finally:
\begin{eqnarray}\label{4.12b}
\nonumber \V & = & -{1 \over 2} \left(R^{(2)} + 2 {Q^2 \over \rho^4}
\right)\\
& = & 2 \left(\frac{Q^2 (r^2-a^2 \cos^2 \theta) - Mr  (r^2-3a^2
\cos^2
\theta)}{\rho^6} \right),
\end{eqnarray}
where we have used the fact that $R^{(2)} =  -F^{\prime \prime}$.

It remains to determine the normal fundamental form $\mu_{RSA}$. Now
as
$\mu_{RSA}=\mu_{[RS]A}$, we can write $\mu_{RSA} = \mu_A
\epsilon_{RS}$. It is
then straightforward to verify that under the gauge transformation
(6.6)
$\mu_{RSA}$ transforms as:
\begin{equation}\label{4.12c}
\mu_{RSA} \mapsto \tilde{\mu}_{RSA} = \mu_{RSA}+ \epsilon_{RS}
x^\mu_{,A}
\partial_\mu \Psi,
\end{equation}
or in light of the previous definition:
\begin{equation}\label{4.12d}
\mu_{A} \mapsto \tilde{\mu}_{A} = \mu_{A}+ x^\mu_{,A} \partial_\mu
\Psi.
\end{equation}

We define $n_R$ over $\Sigma_\pm$ by parallel transport along a
principal null
trajectory and then by Lie transport along trajectories of the
Killing vector,
effectively fixing a gauge. That is on $\Sigma_\pm:$
\begin{equation}\label{4.13}
l_{\pm}^{\mu} n_R^\nu\;_{;\mu}  =  0,\;\;\;\;\;\;
\xi_{\pm}^{\mu} n_R^\nu\;_{;\mu}  =   n_R^\mu \xi_{\nu ;\mu}.
\end{equation}
With this covariantly constant definition of $n_R$, using equation
(\ref{B1})
in Appendix B, we find that:
\begin{eqnarray}\label{4.14}
\mu_{RS1} & = &  n_R^\mu l_{\pm}^{\nu} n_{S \mu ; \nu} = 0, \\
\nonumber \mu_{RS0} & = &   n_R^\mu n_S^\nu \xi_{\mu ; \nu}
= 1/2 \; \epsilon_{RS} (n_2^\mu n_3^\nu - n_3^\mu n_2^\nu) \xi_{\mu ;
\nu}
 =  i \epsilon_{RS} M^\mu \bar{M}^\nu \xi_{\mu ; \nu}.
\end{eqnarray}
In order to take advantage of the decomposition of $\xi_{\mu ; \nu}$
in terms
of the Kinnersley null tetrad (2.10), we note that $M_{\pm}$ and $m$
are
related
by the following null rotation:
\begin{equation}\label{4.15}
M_{\pm}  = m + E l_{\pm},
\end{equation}
where $E= \xi \cdot m$. Thus:
\begin{equation}\label{4.16}
\mu_{RS0} = -\mu \; \epsilon_{RS},
\end{equation}
where $\mu = - a (1-F) \cos \theta / \rho^2$. If we let $\ell_{\pm
A}=x^\mu_{,A}l_{\pm \mu}$ then
we can write the normal fundamental form in this gauge as:
\begin{equation}\label{4.17}
\mu_{RSA} = \mu \ell_{\pm A} \epsilon_{RS},
\end{equation}
so that here $\mu_A = \mu \; \ell_{\pm A}$.

However, a more convenient choice of gauge has $\mu_{RSA} \propto
\epsilon_{RS}
\eta_A$ where $\eta_A =x^\mu_{,A} \xi_{\mu}$ is a Killing vector on
$\Sigma_{\pm}$, see ref.\cite{FrAl:95}. This corresponds to a choice
of the
function $\Psi$ on $\Sigma$ such that $\eta_A \propto \tilde{\mu}_A =
\mu \;
\ell_{\pm A} + x^\mu_{,A} \partial_\mu \Psi$. If we let $\Psi = \Psi
(r)$, then
it follows that on
$\Sigma:$
\begin{equation}\label{4.18}
x^\mu_{,A} \partial_\mu \Psi = \mp \Psi^{\prime} \left(F \ell_{\pm A}
- \eta_A
\right).
\end{equation}
Clearly, if $\Psi^{\prime} = \pm \mu / F$, then $\tilde{\mu}_A = (\mu
/ F)
\eta_A$.
With this choice of gauge we find that the equations of motion reduce
to:
\begin{equation}\label{4.19}
\left({\,\lower0.9pt\vbox{\hrule \hbox{\vrule height 0.2 cm \hskip
0.2 cm
\vrule height 0.2 cm}\hrule}\,}+ \V + \mu^2/F \right) \tilde{\Phi}_R
+ 2
\frac{\mu}{F}\epsilon_{RS} \eta^A \partial_A \tilde{\Phi}^S = 0,
\end{equation}
where:
\begin{equation}
 \mu  =  -\frac{a(1-F)\cos\theta}{\rho^2},
\end{equation}
\begin{equation}
\V  =  2 \left(\frac{Q^2 (r^2-a^2 \cos^2 \theta) - Mr  (r^2-3a^2
\cos^2
\theta)}{\rho^6} \right).
\end{equation}
Equation (6.21) can also be written in the form:
\begin{equation}
[G^{AB}(\delta_{RT}\nabla_A+\epsilon_{RT}{\cal
A}_A)(\delta_{TS}\nabla_B+\epsilon_{TS}{\cal A}_B)+\delta_{RS}{\cal
V}]\tilde{\Phi}^S=0,
\end{equation}
where ${\cal A}_A\equiv\mu\eta_A/F=(-\mu,\pm\mu/F)$ and we used the
identity
$G^{AB}\nabla_A(\mu\eta_B/F)=0.$ Here ${\cal A}_A$ plays the role of
a vector
potential while ${\cal V}$ is the scalar potential. Notice that the
time
component of ${\cal A}_A$ as well as ${\cal V}$ are finite
everywhere, while
the space component of ${\cal A}_A$ diverges at the static limit
surface. But
this divergence can be removed by a simple world-sheet coordinate
transformation:
\begin{equation}
d\tilde{t}= du_{\pm}\mp F^{-1}(r)dr,\;\;\;\;\;\;\;\;
d\tilde{r}=dr.
\end{equation}
The perturbation equation still takes the form (6.24) but now the
potentials
are given by:
\begin{equation}
\tilde{{\cal A}}_A=(-\mu,0),\;\;\;\;\;\;\;\;\;\;\;\;\tilde{{\cal
V}}={\cal V},
\end{equation}
that is, the potentials $(\tilde{{\cal A}}_A,\;{\cal V})$ are finite
everywhere. There is however a divergence at the static limit surface
in the
time component of $\tilde{{\cal A}}^A,$  but such situations are
well-known
from ordinary electro-magnetism; this divergence does not destroy the
regularity of the solution.

\baselineskip6.8mm
\section{String-Hole Physics}
\hspace{\parindent}

In conclusion we discuss some problems connected with the proposed
string-hole
model  of two-dimensional black and white holes. The basic
observation made in
this paper is that the interaction of a cosmic string with a 4-D
black hole in
which the string is trapped by the 4-D black hole opens new channels
for the
interaction of the black hole with the surrounding matter. The
corresponding
new degrees of freedom are related to excitations of the cosmic
string
(stringons). These degrees of freedom can be identified with physical
fields
propagating in the geometry of the 2-D string hole. There are two
types of
string holes corresponding to two types of the principal Killing
surfaces
$\Sigma_+$ and $\Sigma_-$.  The first of them has the geometry of a
2-D black
hole while the second has the geometry of a 2-D white hole. The
physical
properties of 'black' and 'white' string holes are different. For a
regular
initial state a 'black' string hole at late time is a source of a
steady flux
of thermal 'stringons'.  This effect is an analog of the Hawking
radiation
\cite{haw}.  In the simplest case when a stationary cosmic string is
trapped by
a Schwarzschild black hole, so that the string hole has 2-D
Schwarzschild
metric,  the Hawking radiation of stringons was investigated in
ref.\cite{eric}.   For such string holes their event horizon
coincides with the
event horizon of the 4-D black hole, and the temperature of the
'stringon'
radiation coincides with the Hawking temperature of the 4-D black
hole. For
this reason the thermal excitations of the cosmic string will be in
the state
of thermal equilibrium with the thermal radiation of the 4-D black
hole.

The situation is  different in the general case when a stationary
string is
trapped by a rotating charged black hole.  For the Kerr-Newman black
hole the
static limit surface is located outside the event horizon.  The event
horizon
of the 2-D string hole  does not coincide with the Kerr-Newman black
hole
horizon, except for the case where  the cosmic string  goes along the
symmetry
axis . For this reason the surface gravity, and hence the temperature
of the
2-D black hole differ from the corresponding quantities calculated
for the
Kerr-Newman black hole. The surface gravity of the 2-D black hole is
\begin{equation}
\left.
\kappa^{(2)}=\frac{1}{2}\frac{dF}{dr}\right|_{r=r_{st}}=
\frac{\sqrt{M^2-Q^2-a^2\cos^2\theta}}{2M^2-Q^2+
2M\sqrt{M^2-Q^2-a^2\cos^2\theta}},
\end{equation}
and it is always larger than the surface gravity of the 4-dimensional
Kerr-Newman black hole, equation (5.7). The reason why the
temperature of a 2-D
black hole differs from the temperature of the 4-dimensional
Kerr-Newman black
hole can be qualitatively explained if we note that for quanta
located on the
string surface (stringons) the angular momentum and energy are
related.

In the general case ($a\neq 0\;$) a principal Killing surface in the
Kerr-Newman spacetime is  not geodesic. This property might have some
interesting physical applications. Consider a black string hole and
choose a
point $p$ inside its events horizon but outside the event horizon of
the
4-dimensional Kerr-Newman black hole.  Consider a timelike line
$\gamma_0$
representing a static observer located outside the horizon of the 2-D
black
hole at $r=r_0$. There evidently exists an ingoing principal null ray
crossing
$\gamma_0$ and passing through $p$. It was shown that there exists a
future-directed 4-D null geodesic which begins at $p$ and crosses
$\gamma_0$.
In other words a causal signal from $p$ propagating in the 4-D
embedding
spacetime can connect points of the 2-D string hole interior with its
exterior.
For this reason stringons propagating inside the 2-D string hole can
interact
with the stringons in the 2-D string hole exterior.  Such an
interaction from
the 2-D point of view is acausal. This interaction of Hawking
stringons with
their quantum  correlated partners, created inside the string hole
horizon
might change the spectrum of the Hawking radiation, as well as its
higher
correlation functions. This effect might have an interesting
application for
study of the information loss puzzle.

In conclusion, we have shown that in the case of interaction of a
cosmic string
with a black hole a 2-D string hole can be formed. It opens an
interesting
possibility of testing some of the predictions of 2-D gravity. We do
not know
at the moment whether it is also possible to 'destroy' a 2-D string
hole by
applying physical forces which change its motion and allow the
cosmic string to be extracted back from the ergosphere. We hope to
return to
this and other questions connected with the unusual physics of string
holes
elsewhere.

\section*{Acknowledgements}
\hspace{\parindent}
The authors benefitted from helpful discussions with J.Hartle and
W.Israel. The
work of V.F. and A.L.L  was supported by NSERC, while the work by
S.H. was
supported by the Canadian Commonwealth Scholarship and Fellowship
Program.

\appendix
\baselineskip6.8mm
\section{String Black Holes and Dilaton-Gravity}
\hspace{\parindent}

In this appendix we show that the 2-D string holes, can also be
obtained as
solutions of 2-D dilaton gravity with a suitably chosen dilaton
potential. To
be more specific, we consider the following action of 2-D
dilaton-gravity:
\begin{equation}
{\cal S}=\frac{1}{2\pi}\int dt
dx\;\sqrt{-g}\;e^{-2\phi}\;[R+2(\nabla\phi)^2+V(\phi)],
\end{equation}
where the dilaton potential $V(\phi)$ will be specified later. In 2
dimensions
we can choose the conformal gauge:
\begin{equation}
g_{\mu\nu}=e^{2\rho}\times
{\mbox{diag.}}(-1,\;1),\;\;\;\;\;\;\;\;\;\;\;\;\rho=\rho(t,x),
\end{equation}
so that:
\begin{equation}
R=2 e^{-2\rho}(\rho_{,tt}-\rho_{,xx}).
\end{equation}
The action (A.1) then takes the form:
\begin{equation}
{\cal S}=\frac{1}{\pi}\int dt dx
\;e^{-2\phi}\;[\rho_{,tt}-\rho_{,xx}+\phi^2_{,x}-
\phi^2_{,t}+\frac{1}{2}e^{2\rho}V(\phi)].
\end{equation}
The corresponding field equations read:
\begin{eqnarray}
&\rho_{,xx}-\rho_{,tt}+\phi_{,tt}-\phi_{,xx}+
\phi^2_{,x}-\phi^2_{,t}+\frac{1}{4}e^{2\rho}(V'-2V)=0,&\nonumber\\
&\phi_{,xx}-\phi_{,tt}+2(\phi^2_{,t}-\phi^2_{,x})+\frac{1}{2}e^{2\rho}
V=0,&
\end{eqnarray}
where $V'\equiv dV/d\phi.$ Now consider the special solutions:
\begin{equation}
\rho=\rho(x),\;\;\;\;\;\;\;\;\;\;\;\;\phi=\phi(x)
\end{equation}
and introduce the coordinate $r:$
\begin{equation}
\frac{dr}{F(r)}=dx,\;\;\;\;\;\;\;\;\;\;\;\;e^{2\rho}=F.
\end{equation}
Then the metric (A.2) leads to:
\begin{equation}
dS^2=-F(r)dt^2+F^{-1}(r)dr^2,
\end{equation}
which is precisely the form of our 2-D string holes (4.18), in the
coordinates
defined by:
\begin{equation}
d\tilde{t}= du_{\pm}\mp F^{-1}(r)dr,\;\;\;\;\;\;
\;\;d\tilde{r}=dr.
\end{equation}
It still needs to be shown that (A.6)-(A.7) is actually a solution to
equations
(A.5). The equations reduce to:
\begin{eqnarray}
&\phi_{,rr}-\phi^2_{,r}+\frac{F_{,r}}{F}\phi_{,r}-
\frac{1}{4F}(V'-2V)=
\frac{F_{,rr}}{2F},&\nonumber\\
&\phi_{,rr}+\frac{F_{,r}}{F}\phi_{,r}-
2\phi^2_{,r}+\frac{1}{2F}V=0.&
\end{eqnarray}
It can now be easily verified that both equations are solved by a
"logaritmic
dilaton" provided the dilaton potential takes the form:
\begin{equation}
V(\phi)=[\frac{2}{r^2}(rF)_{,r}]_{|r=e^{-\phi}/\lambda},
\end{equation}
\begin{equation}
\phi=-\log(\lambda r),\;\;\;\;\;\;\;\;\;\;\;\;\lambda=\mbox{const.}
\end{equation}
for an arbitrary function $F(r).$ For our 2-D string  holes, $F(r)$
is given by
equation (2.3). The dilaton potential (A.11) then takes the explicit
form:
\begin{equation}
V(\phi)=2\lambda^2
e^{2\phi}[1-\frac{4Me^{-\phi}/\lambda-Q^2}
{e^{-2\phi}/\lambda^2+a^2-ab}+
\frac{2e^{-\phi}(2Me^{-2\phi}/\lambda^2-
Q^2e^{-\phi}/\lambda)}
{\lambda(e^{-2\phi}/\lambda^2+a^2-ab)^2}].
\end{equation}
This result holds for the general cone strings. A somewhat simpler
expression
is obtained for strings in the equatorial plane:
\begin{equation}
V(\phi)=2\lambda^2
e^{2\phi}[1-Q^2\lambda^2e^{2\phi}];\;\;\;\;\;\;\;\;\;\;\;\;\theta=\pi/
2
\end{equation}

\section{Gauge Invariance of the Scalar Potential}
\label{B}
\setcounter{equation}{0}
\renewcommand{\theequation}{B.\arabic{equation}}
\hspace{\parindent}

In this appendix we show that $\V_{RS}$, as defined in equation
(\ref{4.6}), is
gauge invariant i.e. invariant under the transformation (\ref{4.7a})
in the
Kerr-Newman spacetime. Let $M = (n_2 + i n_3)/\sqrt{2}$ where $\{n_2,
n_3 \}$
span the two-dimensional vector space normal to the cone string
world-sheet.
Then
under the transformation specified by (\ref{4.7a}) $M^{\mu} \mapsto
\tilde{M}^{\mu}  = e^{i \Psi} M^{\mu} $. We note that the combination
$M^{\mu}
\bar{M}^{\nu}$ is invariant under this transformation.

We will make use of the following equalities:
\begin{eqnarray}\label{B1}
M^{\mu} \bar{M}^{\nu}  & = & 1/2 \;(n_2^{\mu} n_2^{\nu}+ n_3^{\mu}
n_3^{\nu}) -
i/2 \;(n_2^{\mu} n_3^{\nu} - n_3^{\mu} n_2^{\nu}),\\
\label{B2}
M^{\mu} M^{\nu} & = & 1/2 \; (n_2^{\mu} n_2^{\nu} - n_3^{\mu}
n_3^{\nu}) + i/2
\; (n_2^{\mu} n_3^{\nu} + n_3^{\mu} n_2^{\nu}).
\end{eqnarray}
Now consider:
\begin{eqnarray}\label{B3}
\nonumber G^{AB}x^\mu_{,A}x^\nu_{,B}R_{\mu\rho\sigma\nu}n^\rho_R
n^\sigma_S
& = & (g^{\mu \nu} - \delta^{TQ} n_T^{\mu} n_Q^{\nu})
R_{\mu\rho\sigma\nu}n^\rho_R n^\sigma_S \\
& = & - R_{\rho \sigma} n^\rho_R n^\sigma_S - \delta^{TQ}
R_{\mu\rho\sigma\nu}
n_T^{\mu} n_Q^{\nu} n^\rho_R n^\sigma_S.
\end{eqnarray}
The second term on the right hand side can be written as:
\begin{eqnarray}\label{B4}
\nonumber \delta^{TQ} R_{\mu\rho\sigma\nu} n_T^{\mu} n_Q^{\nu}
n^\rho_R
n^\sigma_S & = &
(n_2^{\mu} n_2^{\nu}+ n_3^{\mu} n_3^{\nu})R_{\mu\rho\sigma\nu}
n^\rho_R
n^\sigma_S \\
\nonumber & = &  \delta^{RS} R_{\mu\rho\sigma\nu} n_2^{\mu} n_2^{\nu}
n^\rho_3
n^\sigma_3 \\
& = & - \delta^{RS} R_{\mu\rho\sigma\nu} M^{\mu} M^{\nu} \bar{M}^\rho
\bar{M}^\sigma,
\end{eqnarray}
making use of (\ref{B1}) and the symmetries of the Riemann tensor
only. This
form is explicitly gauge invariant in any spacetime geometry.

It remains to verify that the term $R_{\rho \sigma} n^\rho_R
n^\sigma_S$ is
also gauge invariant. We note that $M$ and the complex null vector
$m$ of
the Kinnersley tetrad are related by the null rotation $M=m +
E\;l\;.$ We may
then use the fact that $m$ and $l_{\pm}$ are eigenvectors of $R_{\rho
\sigma}$
(see equation (6.8)) to show:
\begin{equation}
R_{\rho \sigma} M^\rho M^\sigma = R_{\rho \sigma} \left( m^\rho
m^\sigma + 2 E
m^\rho l^\sigma_{\pm} + E^2 l^\rho_{\pm} l^\sigma_{\pm}\right) = 0,
\end{equation}
Notice that this holds in any gauge as $M^\rho M^\sigma  \mapsto
\tilde{M}^\rho
\tilde{M}^\sigma = e^{2i \Psi} M^\rho M^\sigma$.  Thus equating real
and
imaginary parts of $R_{\rho \sigma} M^\rho M^\sigma $ to zero one
finds:
\begin{equation}\label{B5}
R_{\rho \sigma} n^\rho_2 n^\sigma_2 = R_{\rho \sigma} n^\rho_3
n^\sigma_3,
\;\;\;\;\; R_{\rho \sigma} n^\rho_2 n^\sigma_3 = - R_{\rho \sigma}
n^\rho_3
n^\sigma_2 = 0.
\end{equation}
Thus under a gauge transformation, we find that:
\begin{eqnarray}\label{B6}
\nonumber R_{\rho \sigma} \tilde{n}^\rho_2 \tilde{n}^\sigma_3 & = &
R_{\rho
\sigma} (\cos \Psi n^\rho_2 -  \sin \Psi n^\rho_3)(\sin \Psi
n^\sigma_2 +  \cos
\Psi n^\sigma_3) \\
& = & 0.
\end{eqnarray}
It then follows that:
\begin{eqnarray}\label{B7}
\nonumber R_{\rho \sigma} \tilde{n}^\rho_2 \tilde{n}^\sigma_2 & = &
R_{\rho
\sigma} (\cos \Psi n^\rho_2 -  \sin \Psi n^\rho_3)(\cos \Psi
n^\sigma_2 - \sin
\Psi n^\sigma_3) \\
& = & R_{\rho \sigma} n^\rho_2 n^\sigma_2.
\end{eqnarray}
Similarly $R_{\rho \sigma} n^\rho_3 n^\sigma_3$ remains unchanged
under
rotation.
Thus we conclude that  $\V_{RS}$ is gauge invariant as
$\Omega_{RAB}\Omega_S\hspace*{1mm}^{AB}$ vanishes independently of
gauge in the
Kerr-Newman spacetime.


\begin{thebibliography}{11}
\bibitem{brown}J.D.Brown, Lower Dimensional Gravity (World
Scientific,
Singapore, 1988).
\bibitem{Witt:91} E.Witten, Phys. Rev., {\bf D44}, 314 (1991).
\bibitem{MaSeWa:91} G.Mandal, A.M.Sengupta, and S.R.Wadia,
Mod. Phys. Lett., {\bf 6}, 1685 (1991).
\bibitem{mann}R.B.Mann, Lower Dimensional Black Holes: Inside and
out, Talk
given at "Heat Kernels and Quantum Gravity", Winnipeg, Canada, Aug
1994.
gr-qc/9501038.
\bibitem{moss}S.Lonsdale and I.Moss, Nucl. Phys., {\bf B298}, 693
(1988).
\bibitem{FroZel:89} V.Frolov, V. Skarzhinski, A. Zelnikov and O.
Heinrich,
Phys. Lett., {\bf B224}, 255 (1989).
\bibitem{FrAl:95}V.P.Frolov and A.L.Larsen, Nucl. Phys., {\bf B449},
149
(1995).
\bibitem{FrAl:94}A.L.Larsen and V.Frolov, Nucl. Phys., {\bf B414} 129
(1994).
\bibitem{guv}J.Guven, Phys. Rev., {\bf D8}, 5562 (1993).
\bibitem{car}B.Carter, Phys. Rev., {\bf D48}, 4835 (1993)
\bibitem{vil}J.Garriga and A.Vilenkin, Phys. Rev., {\bf D44} 1007
(1991).
\bibitem{boy}R.H.Boyer and R.W.Lindquist, J. Math. Phys., {\bf 8},
265 (1967).
\bibitem{GoSa:62} J.N.Goldberg and R.K.Sachs, Acta Phys. Polon.
Suppl., {\bf
22}
13 (1962).
\bibitem{vil2}A.Vilenkin, Phys. Rep., {\bf 121}, 263 (1985).
\bibitem{shell}E.P.S.Shellard and A.Vilenkin, Cosmic Strings
(Cambridge
University Press, Cambridge, 1994).
\bibitem{mtw}C.W.Misner, K.S.Thorne and J.A.Wheeler, Gravitation
(W.H.Freeman,
San Francisco, 1973).
\bibitem{eis}L.P.Eisenhart, Riemannian Geometry (Princeton University
Press,
Princeton NJ, 1964).
\bibitem{haw} S.W.Hawking, Comm. Math. Phys., {\bf 43}, 199 (1975).
\bibitem{eric}A.Lawrence and E.Martinec, Phys. Rev., {\bf D50}, 2680
(1994).
\end{thebibliography}
\end{document}